\documentclass[%
reprint,
superscriptaddress,
nofootinbib,
nobibnotes,
 amsmath,amssymb,
 aps,
prl
]{revtex4-2}

\usepackage{graphicx}
\usepackage{dcolumn}
\usepackage{bm}
\usepackage{orcidlink}
\makeatletter
\let\old@makecaption=\@makecaption
\usepackage{subcaption}
\let\@makecaption=\old@makecaption
\makeatother
\usepackage{amsmath}
\usepackage{tikz}

\newcommand{\subfigpanel}[2]{%
  \begin{tikzpicture}%
    \node[anchor=south west,inner sep=0] (img) at (0,0)%
      {\includegraphics[width=\linewidth]{#2}};%
    \node[anchor=north west,xshift=30pt,yshift=-8pt,%
          fill=white,inner sep=2pt,rounded corners=1pt,%
          font=\scriptsize\bfseries] at (img.north west)%
      {(#1)};%
  \end{tikzpicture}%
}

\newcommand{\subfigpanelwhiteleft}[2]{%
  \begin{tikzpicture}%
    \node[anchor=south west,inner sep=0] (img) at (0,0)%
      {\includegraphics[width=\linewidth]{#2}};%
    \node[anchor=north west,xshift=32pt,yshift=-16pt,%
          font=\scriptsize\bfseries,text=white] at (img.north west)%
      {(#1)};%
  \end{tikzpicture}%
}

\newcommand{\subfigpanelwhite}[2]{%
  \begin{tikzpicture}%
    \node[anchor=south west,inner sep=0] (img) at (0,0)%
      {\includegraphics[width=\linewidth]{#2}};%
    \node[anchor=north west,xshift=165pt,yshift=-16pt,%
          font=\scriptsize\bfseries,text=white] at (img.north west)%
      {(#1)};%
  \end{tikzpicture}%
}

\newcommand{\dd}{\,d}
\newcommand{\Ai}{\operatorname{Ai}}
\newcommand{\Bi}{\operatorname{Bi}}

\newcommand{\im}{\operatorname{Im}}

\begin{document}
\title{Beam-Level Nonlinear Compton Spectra via a Neural Network Surrogate Model}

\author{Antonina Timoshenko}
\affiliation{%
 Skolkovo Institute of Science and Technology, Moscow, Russia
}%
\author{Maxim Malakhov}
\affiliation{%
 Skolkovo Institute of Science and Technology, Moscow, Russia
}%
\affiliation{%
 National Research Nuclear University MEPhI, Moscow, Russia
}%
\author{Alexander Fedotov}
\affiliation{%
 National Research Nuclear University MEPhI, Moscow, Russia
}%
\author{Sergey Rykovanov} %
\affiliation{%
 Skolkovo Institute of Science and Technology, Moscow, Russia
}%

\email{s.rykovanov@skoltech.ru}

\begin{abstract}
Nonlinear Compton scattering enables sources of intense high-energy radiation. However, predicting source-level spectra for realistic electron beams typically requires computationally expensive trajectory-based calculations, whereas existing analytical models for spectral envelopes are limited to a restricted range of laser pulse parameters. Here, we demonstrate that, for electron beams with sufficiently broad phase-space distributions, beam-level nonlinear Compton spectra can be predicted without lengthy numerical calculations. Our approach combines a fast neural-network surrogate for single-electron spectral envelopes with particle-wise Lorentz transformations. Benchmarked against trajectory-based numerical calculations with thousands of macroparticles, the surrogate reproduces the macroscopic spectral-angular structure while reducing the computational cost by orders of magnitude.
\end{abstract}

\maketitle

The development of high-intensity laser systems has opened up new possibilities for creating compact ultra-high-brightness gamma-ray sources based on nonlinear Compton scattering \cite{Nedorezov:PhysUsp2004, Nedorezov:2021, HIGS}. Unlike the linear regime, the interaction of relativistic electrons with intense laser fields involves the absorption of multiple photons, leading to a complex harmonic structure in the radiation spectrum \cite{Sarachik, seipt2016analytical, Krajewska:PRA2012}. For laser pulses with realistic temporal envelopes, such as Gaussian, the inhomogeneous ponderomotive shift results in significant spectral broadening, which often masks individual harmonics and reduces the peak brilliance of the source \cite{timoshenko2025pulse, Kharin:PRA2016, Krafft:PRL2004, Rykovanov2014Thomson}.

Accurate prediction of source-level spectral-angular emission is essential for the design and optimization of next-generation laser-electron sources. Typically, this is achieved through Monte Carlo methods \cite{sun2011theoretical, ridgers2014modelling, blackburn2014quantum}, differential cross-section methods \cite{pratt2010compton, bamber1999studies, seipt2011nonlinear, Rykovanov2014Thomson}, and methods based on calculating electron trajectories \cite{10.1007/978-3-032-22051-6_26, thomas2010algorithm, malakhov2025calculation}. However, realistic beam calculations require evaluating the radiation of thousands of electrons with different energies and transverse momenta. Resolving the rapidly oscillating finite-pulse spectra of every particle makes such calculations computationally expensive and limits their use in parameter scans and source optimization. Analytical envelope models offer a faster alternative, but their applicability is restricted to specific pulse regimes \cite{nikishov1964quantum, heinzl2020locally, Kharin:PRA2016}.

To address this challenge, we exploit a physical property of electron beams with sufficiently broad phase-space distributions. Energy spread and transverse momentum spread dephase the finite-pulse interference fringes present in the spectra of individual electrons. Upon averaging over the beam phase space, these fringes are washed out, so that the beam-level spectral-angular emission can be approximated by an incoherent sum of smooth single-electron spectral envelopes. This avoids resolving the highly oscillatory spectrum of every electron individually.

To evaluate these envelopes rapidly, we develop NCS-Net, a neural-network surrogate for the first-harmonic single-electron spectral envelope. The model takes the normalized laser amplitude $a_0$\footnote{The dimensionless amplitude of the laser pulse is $a_0=eE/mc\omega_L$, where $E$ is the electric field amplitude, $\omega_L$ is the laser pulse frequency, $e$ and $m$ are the electron charge magnitude and mass, respectively, and $c$ is the speed of light in vacuum. We use dimensionless variables $t \rightarrow \omega_L t$, $\omega \rightarrow \omega/\omega_L$, and $z \rightarrow z\omega_L/c$.}, the dimensionless duration $\tau$ of a Gaussian laser pulse, and the observation angle $\theta$ in the electron rest frame, and predicts the corresponding smooth spectral profile. We also develop an analytical model of the nonlinear Compton spectral envelope that provides physically consistent training data in the long-pulse nonlinear regime (Appendix~\ref{appx:formulas}). By targeting the smooth envelope rather than the oscillatory fine structure of the full spectrum, NCS-Net retains the total photon yield, spectral width, peak amplitude, and peak position relevant to beam-level emission. The analytical model removes unphysical singularities present in previous approximations \cite{nikishov1964quantum, heinzl2020locally, Kharin:PRA2016} and extends the approach of Ref.~\cite{malakhov2026analyticalcompton}. 

We consider Compton scattering of a circularly polarized Gaussian laser pulse in the initial electron rest frame; the resulting spectra can be Lorentz-transformed to any reference frame~\cite{Sarachik}. The NCS-Net covers both linear ($a_0 \ll 1$) and mildly nonlinear ($a_0 \gtrsim 1$) Compton scattering regimes. For $a_0^2 \tau \gg 1, \tau \gg 1$, the training dataset is generated by analytical formulas (Appendix~\ref{appx:formulas}). Otherwise, the spectrum is calculated numerically via the quadratic-interpolation method \cite{thomas2010algorithm,Chen2013PRSTAB,malakhov2025calculation}. For off-axis angles, the network predicts the first harmonic. The training and test datasets cover $a_0 \in [0.01,\,2]$, $\tau \in [20\pi,\,200\pi]$, and $\theta \in [\pi/2,\,\pi]$.

\begin{figure*}[t]
    \centering
    \includegraphics[width=1.0\textwidth]{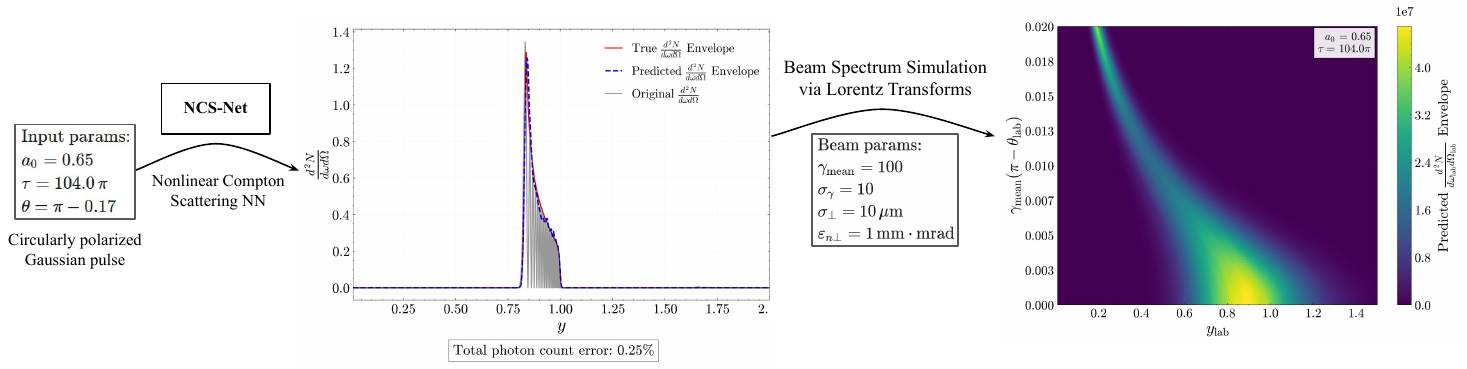}
    \caption{Principle of operation of NCS-Net. Given the laser amplitude $a_0$, pulse duration $\tau$, and polar observation angle $\theta$, the surrogate network predicts the first-harmonic spectral envelope. The spectra are plotted against the Doppler-normalized frequency $y=\omega(1-\beta)/[\omega_L(1+\beta)]$, where $\beta$ is the normalized velocity of an electron counterpropagating with respect to the laser pulse along the $z$ axis.}
    \label{fig:NN_idea}
\end{figure*}

The developed NCS-Net surrogate model takes a low-dimensional parameter vector ($a_0, \tau, \theta$) and transforms it into a high-resolution spectral distribution (Fig.~\ref{fig:NN_idea}). The network architecture, training procedure, and loss function are described in Appendix~\ref{appx:nn_arch}.

Figure~\ref{fig:1D} compares NCS-Net predictions with reference spectral lineouts from the held-out test set. Each curve represents the differential photon yield $d^2 N/d\omega d\Omega$ as a function of frequency at a fixed polar observation angle $\theta$. Owing to the axial symmetry of the circularly polarized laser pulse, the spectrum is independent of the azimuthal angle $\varphi$. The reference spectra are obtained using the analytical or numerical procedures described above, depending on the corresponding parameter range.
\begin{figure*}[t!]
  \centering
  \begin{subfigure}[b]{0.45\textwidth}
    \subfigpanel{a}{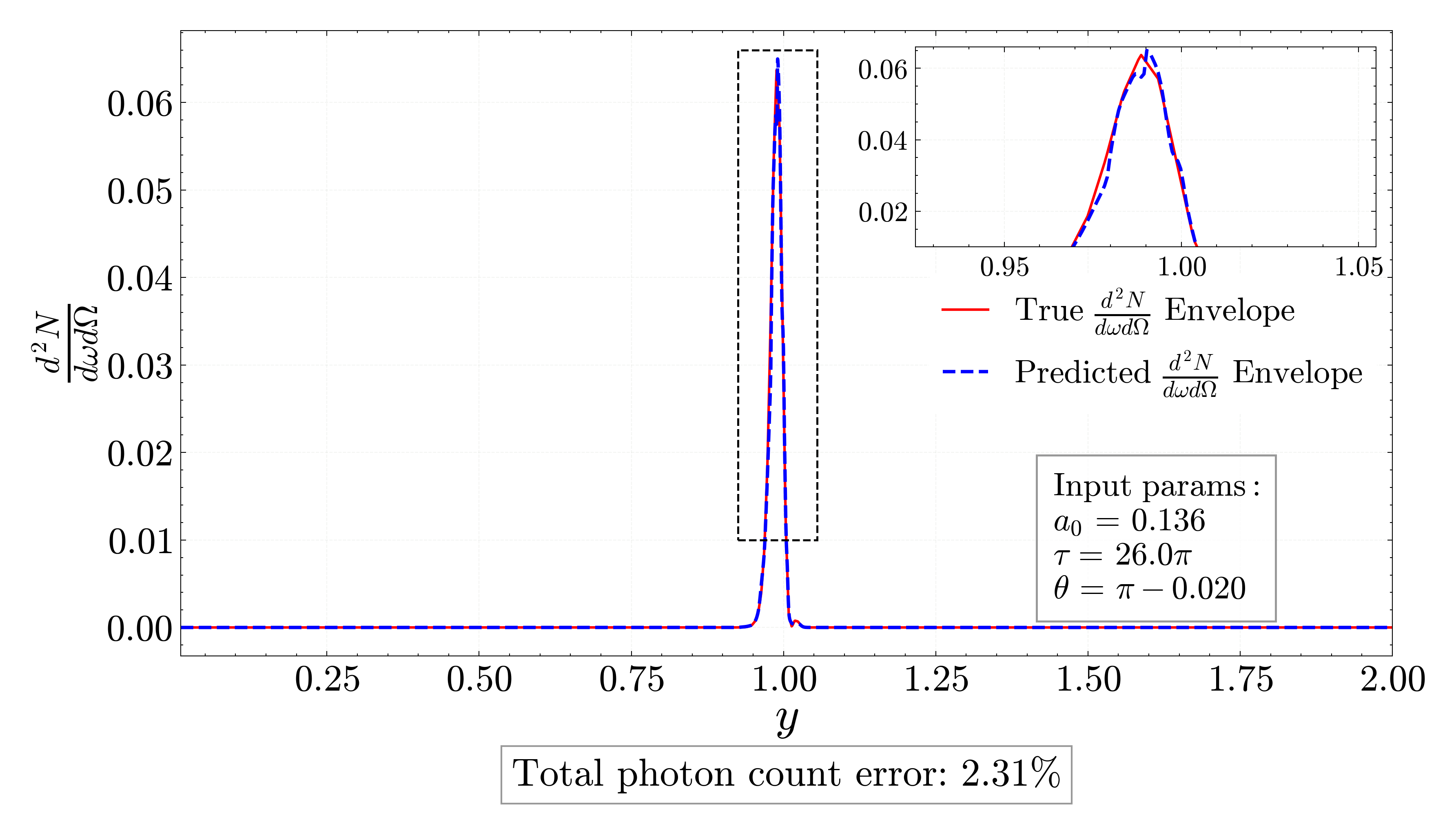}
    \phantomcaption
    \label{fig:img1}
  \end{subfigure}
  \hfill
  \begin{subfigure}[b]{0.45\textwidth}
    \subfigpanel{b}{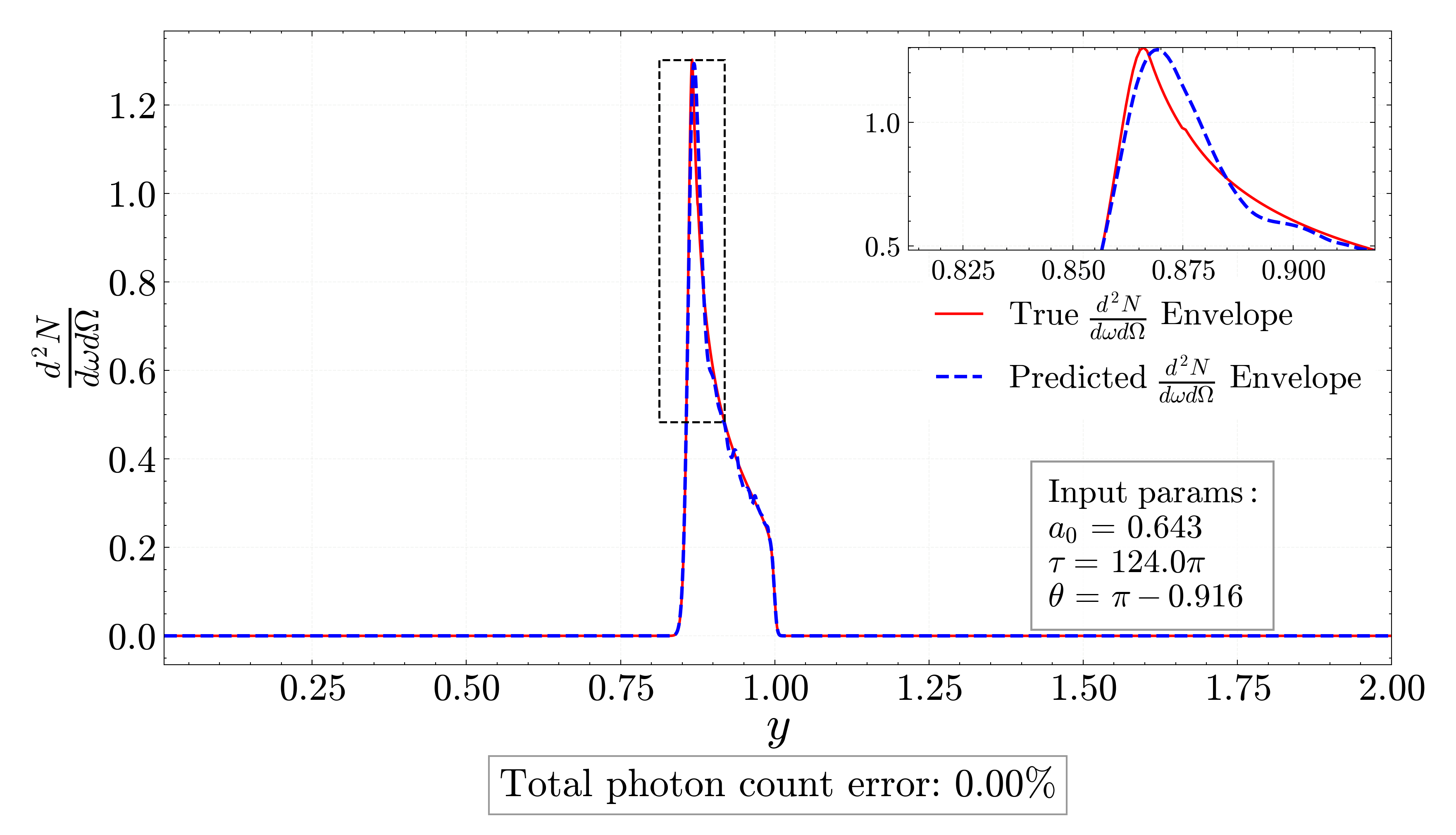}
    \phantomcaption
    \label{fig:img2}
  \end{subfigure}
  \begin{subfigure}[b]{0.45\textwidth}
    \subfigpanel{c}{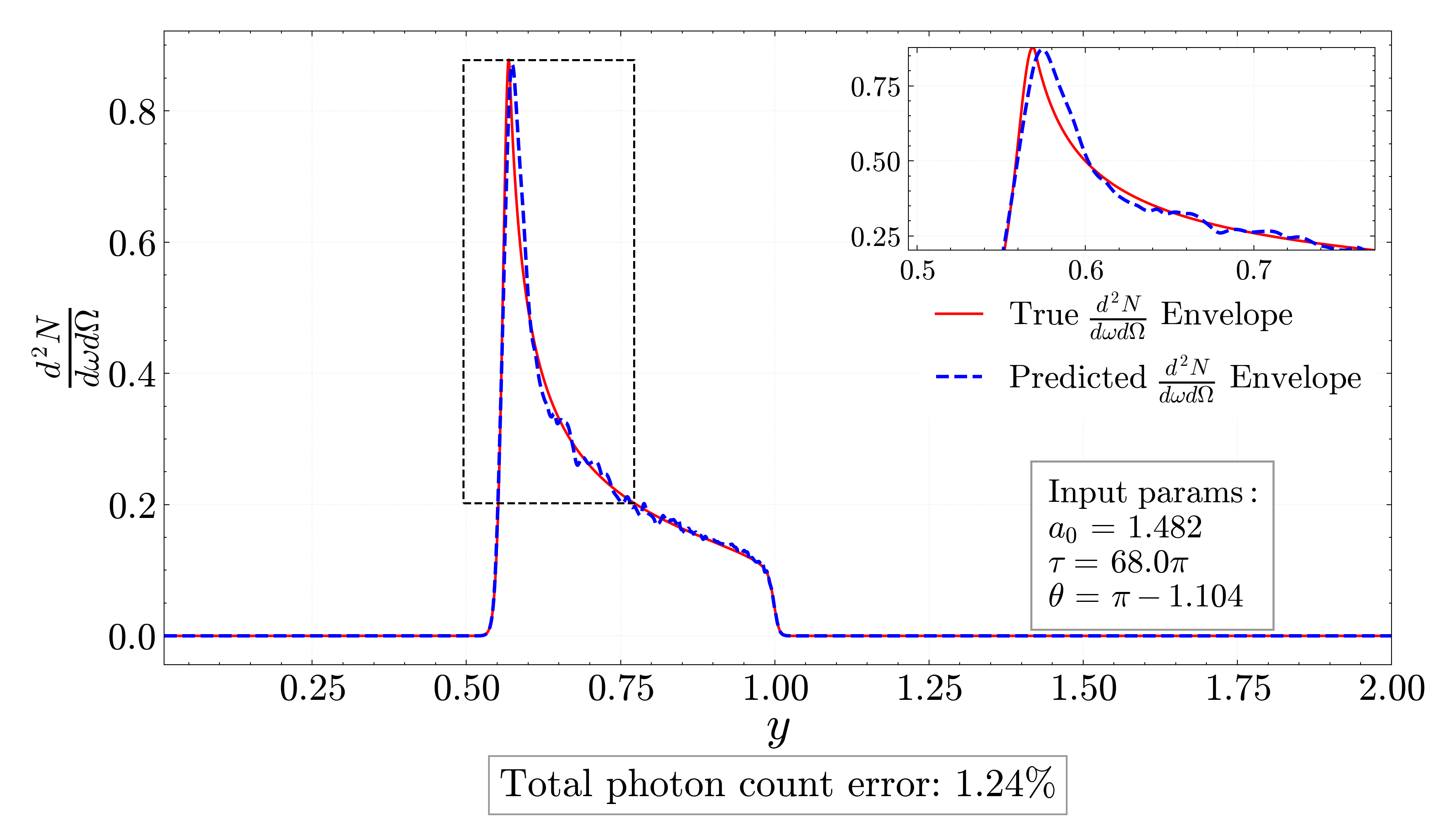}
    \phantomcaption
    \label{fig:img3}
  \end{subfigure}
  \hfill
  \begin{subfigure}[b]{0.45\textwidth}
    \subfigpanel{d}{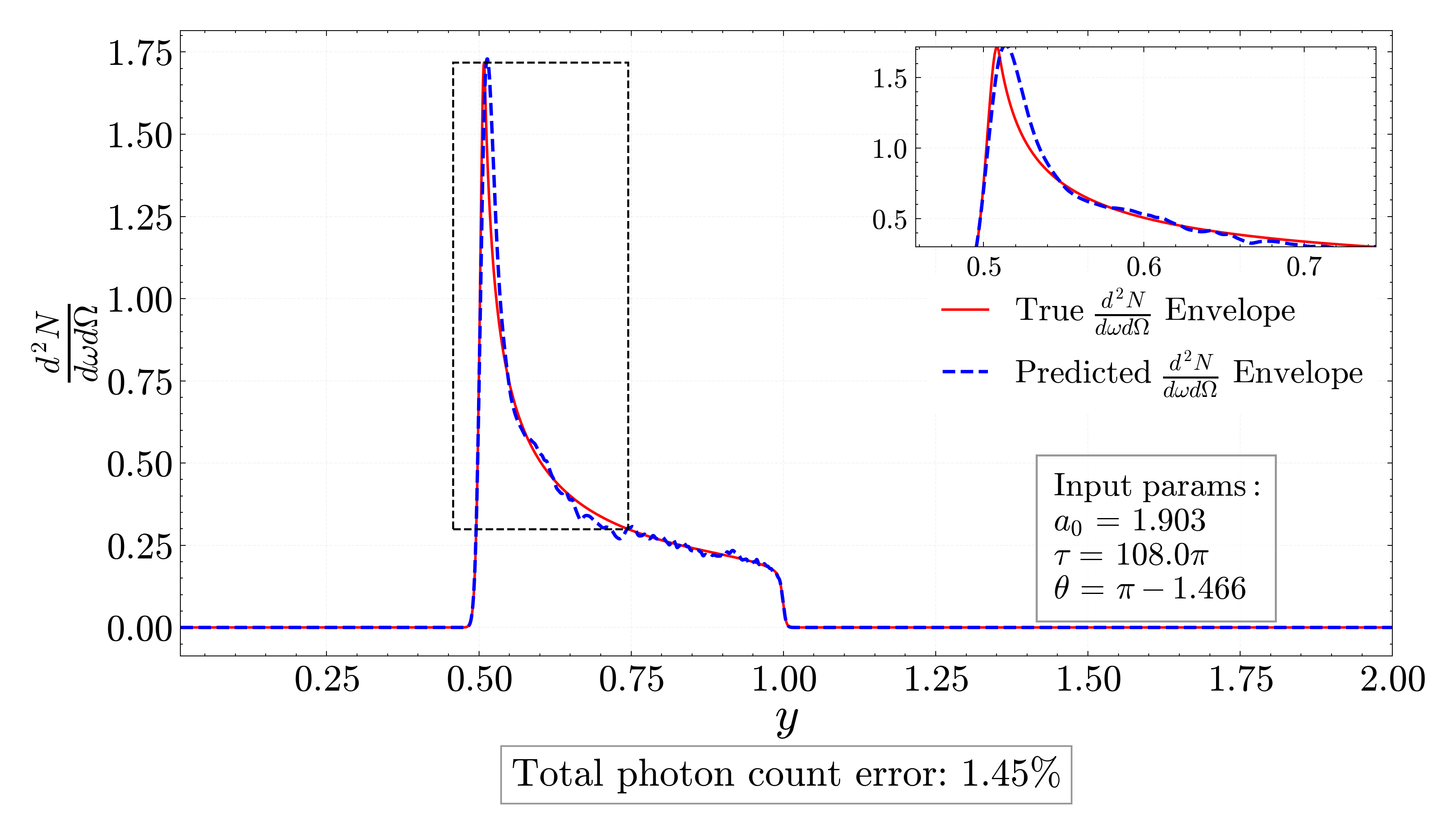}
    \phantomcaption
    \label{fig:img4}
  \end{subfigure}
  \caption{NCS-Net inference results for representative test samples. Each curve shows the spectral lineout $\dd^2N/\dd\omega,\dd\Omega$ at a fixed polar observation angle $\theta$; owing to axial symmetry, the result is independent of the azimuthal angle $\varphi$. The red solid line denotes the reference spectral envelope obtained from the analytical model (Appendix~\ref{appx:formulas}), and the blue dotted line denotes the NCS-Net prediction.}
  \label{fig:1D}
\end{figure*}

The NCS-Net accurately reproduces the smooth spectral envelope, including the peak position, peak amplitude, and overall spectral width. To quantify the agreement, we integrate each spectral lineout over frequency and compare the resulting photon yield per unit solid angle, $\dd N/\dd\Omega$, with the reference value. The average relative error of this quantity over the test set does not exceed 2.5\%.

\begin{figure*}[t!]
  \centering
  \begin{subfigure}[b]{0.45\textwidth}
    \subfigpanelwhite{a}{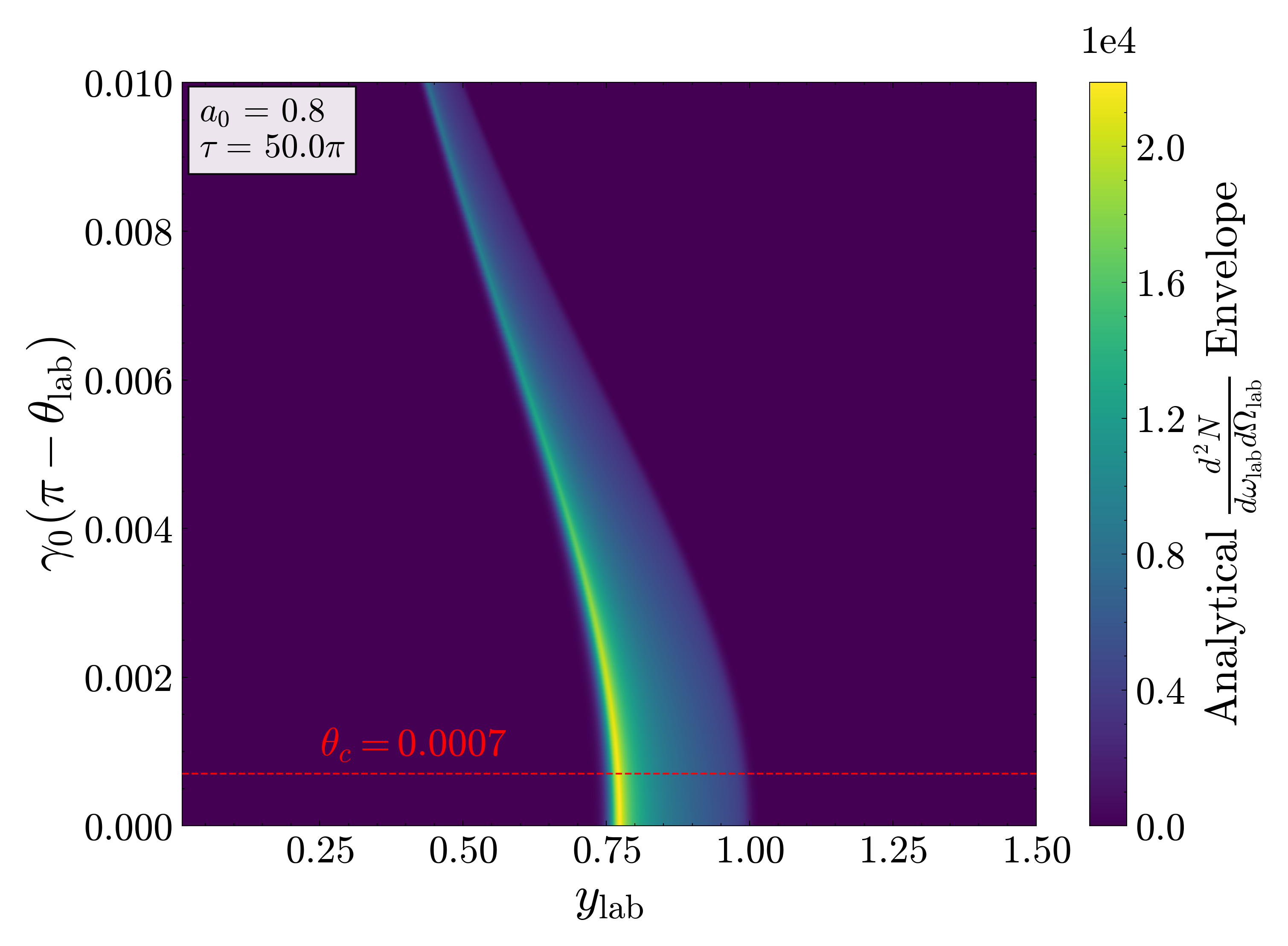}
    \phantomcaption
    \label{fig:2D_img3}
  \end{subfigure}
  \hfill
  \begin{subfigure}[b]{0.45\textwidth}
    \subfigpanelwhite{b}{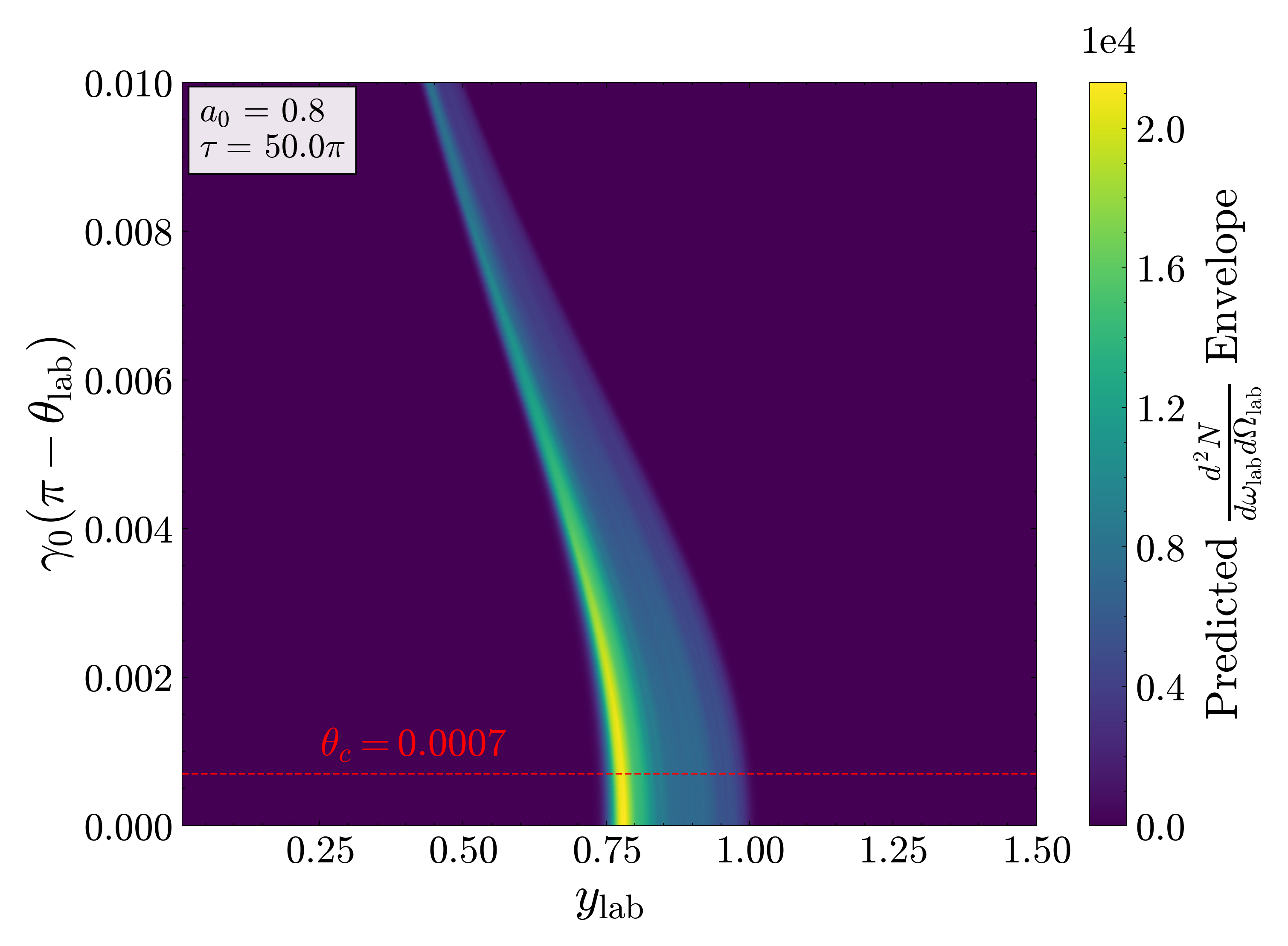}
    \phantomcaption
    \label{fig:2D_img4}
  \end{subfigure}
  \caption{Two-dimensional frequency-angular spectral distributions: (a) from the analytical model (Appendix~\ref{appx:formulas}); (b) from NCS-Net. Spectral distributions are Lorentz-transformed to the laboratory frame with initial electron energy $\gamma_0 = 100$.}
  \label{fig:2D}
\end{figure*}

Beyond one-dimensional envelope predictions, the developed neural network enables rapid reconstruction of two-dimensional frequency-angular distributions (Fig.~\ref{fig:2D}) and, consequently, the photon yield after collimation. For fixed laser pulse parameters ($a_0, \tau$), the network is evaluated over a dense grid of scattering angles $\theta$, and the resulting one-dimensional spectral envelopes are stitched into a continuous frequency-angular map. Owing to the axial symmetry of the circularly polarized pulse, the distribution is independent of the azimuthal angle. The photon yield within a collimation angle $\theta_c$ is then obtained by integrating the reconstructed map over frequency and the corresponding solid-angle acceptance.

A comparison with the analytical reference calculation demonstrates accurate reproduction of the harmonic-band boundaries and of the angular dependence of the spectral-peak frequency. Figures \ref{fig:2D_img3}, \ref{fig:2D_img4} show the frequency-angular distributions of the differential photon yield, Lorentz-transformed into the laboratory reference frame for an electron with initial Lorentz factor $\gamma_0 = 100$. Averaged over the test dataset, the relative error in the total number of photons within the collimation angle $\theta_c$ does not exceed 1.0\%. 

\begin{figure*}[t!]
  \centering
  \begin{subfigure}[b]{0.46\textwidth}
    \subfigpanelwhiteleft{a}{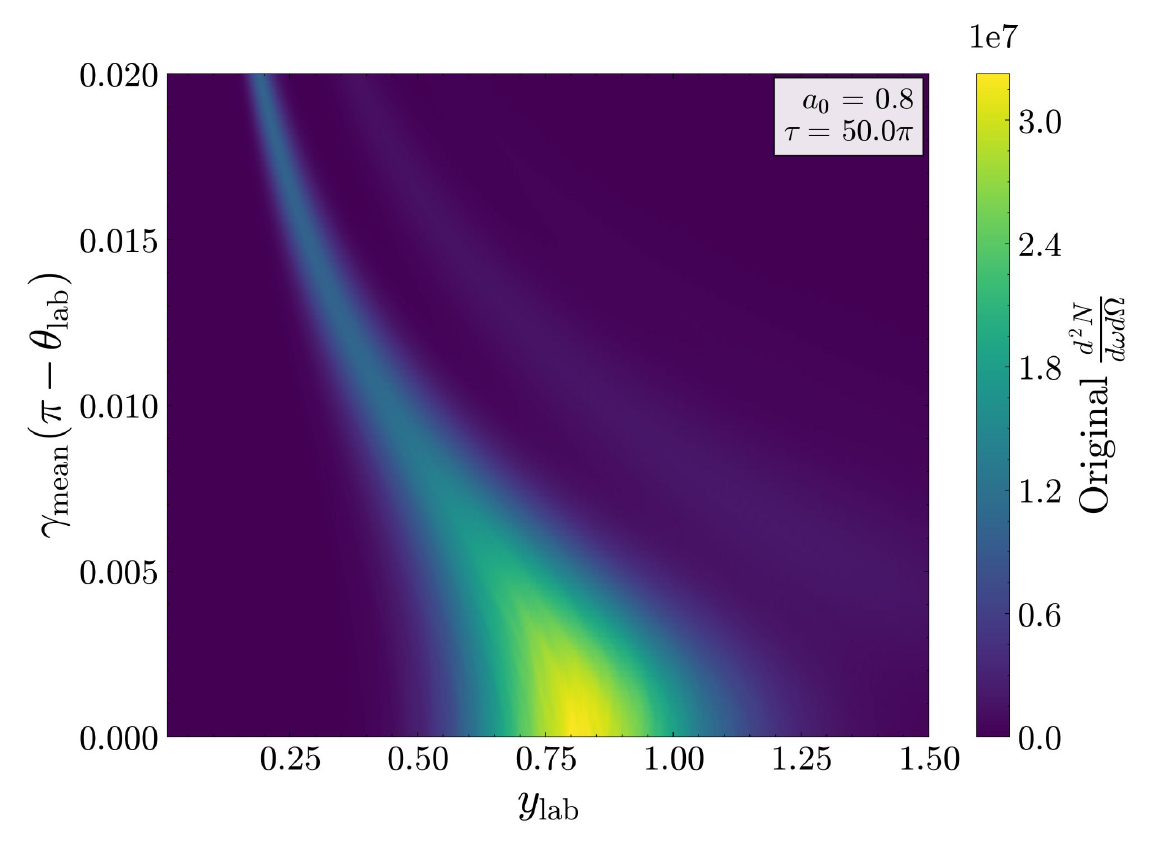}
    \phantomcaption
    \label{fig:2D_beam_img1}
  \end{subfigure}
  \hfill
  \begin{subfigure}[b]{0.45\textwidth}
    \subfigpanelwhiteleft{b}{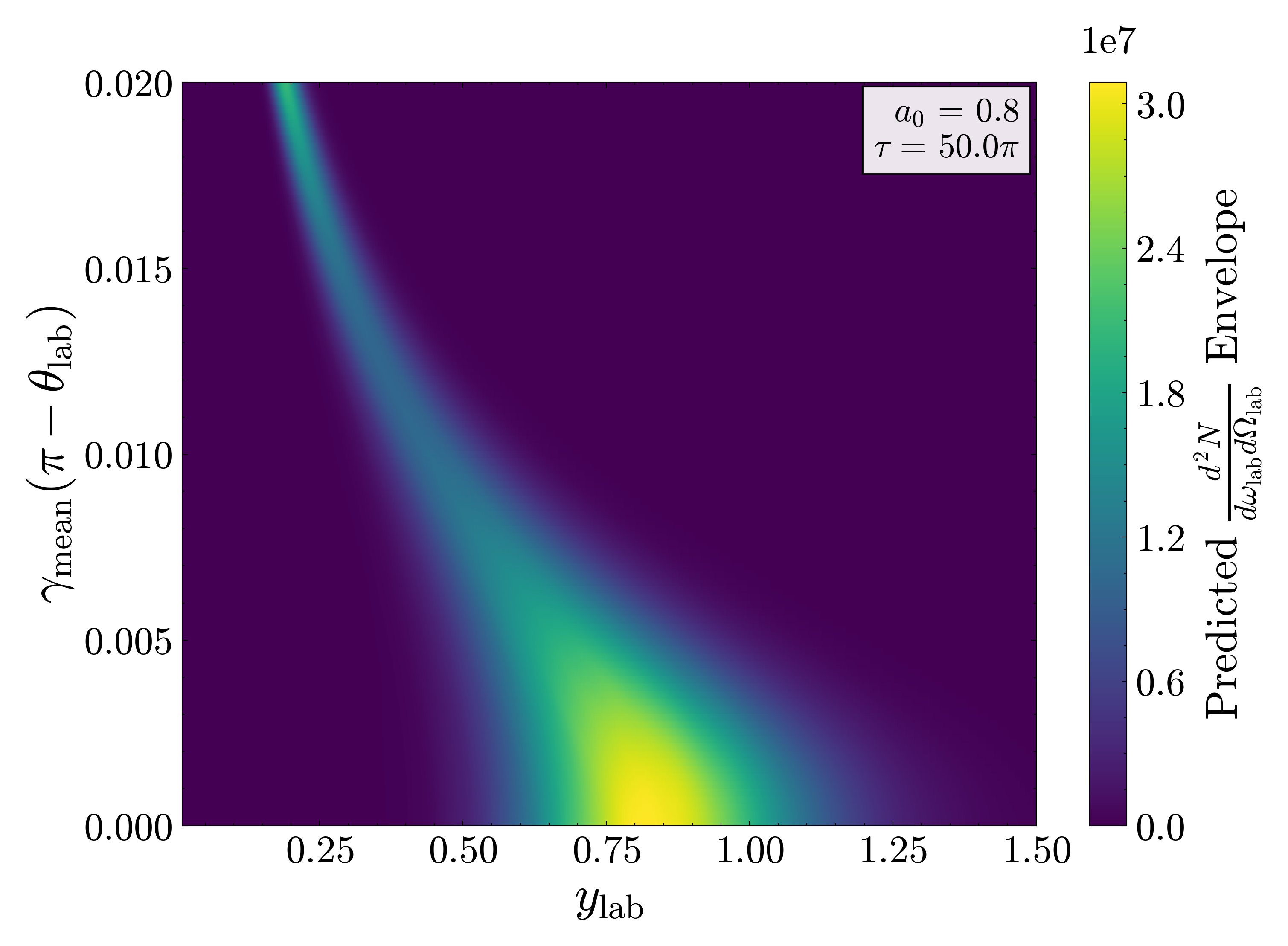}
    \phantomcaption
    \label{fig:2D_beam_img2}
  \end{subfigure}
  \caption{Two-dimensional frequency-angular spectral distributions of the electron beam ($10^4$ sampled particles): (a) from the numerical method based on quadratic approximations; (b) from NCS-Net (divided by 2 for comparison with the exact spectrum). The electron beam has the following parameters: $\gamma_{\text{mean}} \approx 100,\; \sigma_{\gamma}\approx 10,\; \text{normalized emittance } \varepsilon_{n\perp} = 1 \text{mm}\cdot \text{mrad},\; \text{transverse dimensions of the beam } \sigma_{\perp}=10 \mu \text{m},\; \text{mean transverse momenta } \mu_{p_{\perp}} = 0$.}
  
  \label{fig:2D_beam}
\end{figure*}

The main result is that the proposed neural network enables rapid construction of the beam-averaged first-harmonic radiation spectrum (Fig.~\ref{fig:2D_beam}). Each particle sampled from the beam distribution is Lorentz-transformed to its rest frame. After predicting the single-electron spectral envelope with NCS-Net, the corresponding spectrum is transformed back to the laboratory reference frame, and the particle contributions are co-added after interpolation onto a common frequency--angle grid.

The exact single-electron spectra contain finite-pulse interference fringes that are not retained in the NCS-Net output. For an electron beam with finite energy spread and emittance, these fringes are dephased during the incoherent summation over particles. In the rapidly oscillating spectral region near the linear edge, the beam-averaged exact spectrum therefore approaches approximately one half of the smooth upper envelope predicted by NCS-Net. Although this is not a sharp criterion, the phase-averaging approximation is expected to be most reliable when the inhomogeneous broadening induced by the beam energy spread and angular divergence is comparable to, or exceeds, the intrinsic nonlinear bandwidth of the single-electron harmonic. Under these conditions, the variation of the fringe pattern across the ensemble is sufficient to wash out the finite-pulse modulation. Accordingly, the NCS-Net result in Fig.~\ref{fig:2D_beam_img2} is rescaled by a factor of $1/2$ before comparison with the numerical reference.

This phase-averaging correction is not expected to be uniform across the entire spectrum. Its accuracy decreases near the nonlinear edge, where the spectral structure is governed by the transition region of the finite-pulse asymptotics rather than by rapidly oscillating interference fringes. The resulting mismatch contributes to the residual discrepancy between the predicted and exact beam spectra. Compared with the exact beam spectrum obtained using the numerical quadratic-interpolation method based on calculating electron trajectories, the NCS-Net prediction reproduces the beam-level frequency--angle distribution with a total photon-count error of about 10\%.

For the broad beam distributions considered here, the energy spread and transverse phase-space spread produce appreciable variations of the single-electron fringe pattern across the ensemble. In the representative cases with $\sigma_{\gamma}/\gamma_{\text{mean}} > 5\%$ and $\varepsilon_{n\perp}/\sigma_{\perp} \geq 10\%$, these variations are sufficient to wash out the finite-pulse interference structure after beam averaging. More generally, the phase-averaging approximation is expected to apply when the beam-induced displacement of the single-electron spectral features across the ensemble is comparable to, or exceeds, the characteristic spacing of the interference fringes. In the present simulations, more than $10^3$ sampled particles are sufficient to converge this ensemble average. Under these conditions, the trajectory-resolved beam spectrum can be approximated by an incoherent sum of spectral-envelope predictions, particularly in regimes where beam-induced broadening dominates the residual finite-pulse modulation.

In addition, the key advantage of the NCS-Net surrogate model is the substantial reduction in computational time relative to direct numerical calculations. The average NCS-Net prediction time for a single 1D spectrum (Fig.~\ref{fig:1D}) on an Intel(R) Core(TM) i9-14900K CPU (24 physical cores, 32 threads) does not exceed $0.3$ms, whereas a numerical calculation using the quadratic-interpolation method on the same CPU and frequency grid requires nearly $350$ms.

Predicting the complete two-dimensional frequency--angular distribution of the beam-level spectrum with $10^4$ macroparticles using NCS-Net (Fig.~\ref{fig:2D_beam_img2}) takes on average $\sim 34$s, including the particle-wise Lorentz transformations. This is approximately 34 times faster than the corresponding beam-level calculation based on the analytical formulas (Appendix~\ref{appx:formulas}). The corresponding numerical calculation using the quadratic-interpolation method (Fig.~\ref{fig:2D_beam_img1}) requires more than $24$h on the same CPU. The proposed surrogate therefore makes rapid parameter scans and source-level optimization feasible at a computational cost that would be prohibitive for trajectory-resolved numerical calculations. Because NCS-Net is implemented in PyTorch, inference over the macroparticle ensemble and the frequency--angle grid can be readily executed on modern GPUs, offering a direct path to substantially faster source-level calculations.

In conclusion, we have demonstrated that, for broad incoherent electron beams in which finite-pulse interference fringes are washed out by beam averaging, beam-level first-harmonic nonlinear Compton spectra can be predicted without trajectory-resolved spectral calculations. We introduced NCS-Net, a fast neural-network surrogate for single-electron spectral envelopes which, combined with particle-wise Lorentz transformations, reconstructs beam-level frequency--angular distributions over the laser and beam parameter ranges considered here. To provide a physically grounded training basis, we additionally developed an analytical model of the nonlinear Compton spectral envelope that extends the approach presented in~\cite{malakhov2026analyticalcompton}.

The proposed framework enables rapid source-level parameter scans and optimization at a computational cost that would be prohibitive for trajectory-resolved simulations. It is particularly relevant to the design of narrow-band gamma-ray sources, where nonlinear Compton effects must be accounted for in the design and optimization of the source parameters \cite{Hartemann2010, NCPM2025}.

\begin{acknowledgments}
Part of the simulations was performed on the Skoltech supercomputer ``Zhores''~\cite{zhores}. The authors thank Dr. I.~Yu. Kostyukov for fruitful discussions. This work was supported by NCPhM Project 6.
\end{acknowledgments}

\providecommand{\noopsort}[1]{}\providecommand{\singleletter}[1]{#1}%
%

\appendix
\makeatletter
\setcounter{secnumdepth}{1}
\makeatother

\section{\label{appx:formulas}Analytical Calculation of the Spectrum of Nonlinear Compton Scattering in Finite-Duration Laser Pulses}

In this Appendix, we collect the analytical formulas used to construct the smooth envelope of the first harmonic in nonlinear Compton scattering from a finite laser pulse. The expressions are based on the asymptotic treatment of the finite-pulse phase integrals developed in \cite{malakhov2026analyticalcompton}.

We work in the rest frame of the initial electron. A circularly polarized laser pulse propagates along the positive $z$ axis and is described by the dimensionless vector potential $\mathbf{a}(\phi)=a_0 g(\phi/\tau)(\mathbf{e}_x\cos\phi+\mathbf{e}_y\sin\phi)/\sqrt{2}$, where $\phi=t-z$ is the laser phase and $\tau$ is the dimensionless pulse duration. The observation direction is specified by the polar angle $\theta$, measured from the positive $z$ axis. The envelope $g$ is chosen in the Gaussian form
\begin{equation}
    g(\phi/\tau)
    =
    \exp\!\left(
        -\frac{\phi^2}{2\tau^2}
    \right).
    \label{eq:app-gaussian-envelope}
\end{equation}
In the asymptotic formulas below we use the rescaled phase $\phi/\tau\to\phi$ and keep the same symbol $\phi$ for brevity. For circular polarization, in the classical limit, the differential number of photons emitted into the solid angle $\dd\Omega$ and frequency interval $\dd\omega$ can be written as
\begin{equation}
  \frac{\dd^2 N}{\dd\omega\,\dd\Omega}
  =
  \frac{\alpha \omega}
       {4\pi^2}
  \left[
      -A_0^2
      +
      \frac{a_0^2}{2}
      \left(
          A_+^2
          +
          A_-^2
          -
          2A_0A_2
      \right)
  \right].
  \label{eq:app-spectrum-general}
\end{equation}
Here and below we use the dimensionless photon frequency
$\omega/\omega_L\to\omega$. For the first harmonic, the phase integrals
entering Eq.~\eqref{eq:app-spectrum-general} are written in the local Bessel
form
\begin{equation}
\begin{pmatrix}
  A_{0}
  \\
  A_{\pm}
  \\
  A_2
\end{pmatrix}
=
\tau D
\begin{pmatrix}
  -g_0^{-1}J_1(\bar\alpha g_0)
  \\
  J_{1\mp1}(\bar\alpha g_0)
  \\
  -g_0J_1(\bar\alpha g_0)
\end{pmatrix}.
\label{eq:app-first-harmonic-integrals}
\end{equation}
Here $g_0=|g(\phi_0)|$, where $\phi_0$ is the saddle point specified below, and $\bar\alpha=a_0\omega\sin\theta/\sqrt{2}$ is the transverse phase parameter. The quantity $D$ contains the oscillating finite-pulse structure.

The form of $D$ depends on the position of the emitted frequency within the
first harmonic. The frequency
$\hat{\omega}_1=1/(1+a_0^2b(\theta))$ is the nonlinear edge of the first
harmonic, while $\omega_0$ is the matching frequency at which we switch from
the Airy approximation to the linear-edge saddle approximation \cite{malakhov2026analyticalcompton}. Thus, for
$\omega\le\hat{\omega}_1$ and $\hat{\omega}_1<\omega<\omega_0$, the quantity
$D$ is represented by the Airy approximation,
\begin{equation}
  D =
  \sqrt{\frac{8\pi^2}{|\zeta''(\phi_0)|}}g_0\Ai(u), 
  \label{eq:app-D-airy}
\end{equation}
where the Airy argument is chosen according to the frequency range,
\begin{equation}
  u=
  \begin{cases}
    \zeta_0,
    & \omega \le \hat{\omega}_1 ,
    \\[1mm]
    -\zeta_0,
    & \hat{\omega}_1<\omega<\omega_0 .
  \end{cases}
  \label{eq:app-y-airy}
\end{equation}
Here $\zeta(\phi)$ denotes the Airy mapping, and its value at the saddle point
is fixed by $\zeta_0=(3\tau|F(\phi_0)|/2)^{2/3}$. The phase entering this
mapping is
\begin{equation}
  F(\phi)=(\omega-1)\phi+\frac{\sqrt{\pi}}{2}\omega a_0^2b(\theta)\textrm{Erf}(\phi),
  \label{eq:app-F}
\end{equation}
where $b(\theta)=\sin^2{\left(\theta/2\right)}/2$. The prefactor in
Eq.~\eqref{eq:app-D-airy} is obtained from the local matching condition
$\zeta''(\phi_0)\zeta^{1/2}(\phi_0)=i\tau F''(\phi_0)$. The saddle points used
in this region are
\begin{equation}
  \phi_0 =
  \begin{cases}
    \displaystyle
    i\sqrt{-\ln{\frac{a_0^2 \omega b(\theta)}{1-\omega}}},
    & \omega \le \hat{\omega}_1 ,
    \\[3mm]
    \displaystyle
    \sqrt{\ln{\frac{a_0^2 \omega b(\theta)}{1-\omega}}},
    & \hat{\omega}_1<\omega<\omega_0  ,
  \end{cases}
  \label{eq:app-phi0-nonlinear}
\end{equation}

For $\omega>\omega_0$, the Airy approximation is replaced by the
envelope-corrected saddle expression. In this region the saddle points are
complex and the corresponding contribution is written as
\begin{multline}
  D =
  \sqrt{\frac{8\pi}{\tau |q''(\phi_0)|}}\,
  |\Sigma|g_0\,
  e^{-\tau \im F(\phi_0)}\\ \times
  \cos\!\left[
      \tau \im q(\phi_0)+\arg(h_0\Sigma)
  \right],
  \label{eq:app-D-linear}
\end{multline}
where $q(\phi)=iF(\phi)-\phi^2/2\tau$ and 
\begin{equation}
  \Sigma
  =
  1+\frac{1}{\tau}
  \left[
  \frac{q^{(4)}(\phi_0)}
       {8[q''(\phi_0)]^2}
  -
  \frac{5[q^{(3)}(\phi_0)]^2}
       {24[q''(\phi_0)]^3}
  \right],
\end{equation}
and $h_0=\sqrt{-2/q''(\phi_0)}$.

The saddle point in the third region is determined by
$\phi_0=\tilde\phi_0+\delta\tilde\phi$, where 
\begin{equation}
  \delta\tilde\phi=-\frac{q'(\tilde\phi_0)}{q''(\tilde\phi_0)} ,
\end{equation}
with
\begin{equation}
  \tilde\phi_0
  =
  \left[\ln{\left(\frac{a_0^2 \omega b(\theta)}{1-\omega}+g^2(\bar{\phi}_0)e^{-(1-\omega)^2}\right)}\right]^{1/2},
\end{equation}
and
\begin{equation}
  \bar\phi_0=\left[\frac{1}{2}W\!\left(-2\tau^2 a_0^4 b^2(\theta)\right)\right]^{1/2},
\end{equation}
where $W$ is the Lambert function.

We now pass from the oscillating expression to the smooth harmonic envelope.
Since the phase integrals enter Eq.~\eqref{eq:app-spectrum-general}
quadratically, the envelope is constructed at the level of $D^2$. The factor
$g_0$ in Eqs.~\eqref{eq:app-D-airy} and \eqref{eq:app-D-linear} is a smooth
local prefactor. We therefore define $D_{\rm env}^2$ as the envelope of
$D^2/g_0^2$, keeping the corresponding factor $g_0^2$ explicitly in the final
spectrum. In the first two regions the oscillations are contained in the Airy
factor. The envelope is chosen to pass through the first maximum of the Airy
function and, after the first zero of $\Bi(u)$, $u_0\approx -1.1737$, the
oscillating Airy tail is replaced by the smooth combination
$\Ai^2(u)+\Bi^2(u)$. In the third region the oscillating factor is the cosine
in Eq.~\eqref{eq:app-D-linear}, so the envelope is obtained by omitting this
factor.
\begin{equation}
  D_{\rm env}^2
  =
  \frac{8\pi^2}{|\zeta''(\phi_0)|}
  \left(
      \Ai^2(u)
      +
      \Bi^2(u)\Theta(u_0-u)
  \right),
  \label{eq:app-Denv-airy}
\end{equation}
where $\Theta$ is the Heaviside step function. For the third region we define
\begin{equation}
  D_{\rm env}^2
  =
  \frac{8\pi}{\tau |q''(\phi_0)|}\,
  |\Sigma|^2
  e^{-2\tau \im F(\phi_0)}.
  \label{eq:app-Denv-linear}
\end{equation}
Thus, in the expression for the harmonic envelope, the squared oscillating
quantity $D^2$ is replaced by its smooth envelope $g_0^2D_{\rm env}^2$.
Substituting Eq.~\eqref{eq:app-first-harmonic-integrals} into
Eq.~\eqref{eq:app-spectrum-general}, one obtains
\begin{equation}
  \frac{\dd^2 N}{\dd\omega\,\dd\Omega}
  =
  \frac{\alpha \omega \tau^2}
       {4\pi^2}
  \left[
      -J_{1}^2
      +
      \frac{a_0^2}{2}g_0^2
      \left(
          J_{0}^2
          +
          J_{2}^2
          -
          2J_{1}^2
      \right)
  \right]D^2_{\rm env},
  \label{eq:app-spectrum-envelope}
\end{equation}
where $J_n\equiv J_n(\bar\alpha g_0)$.

The position of the first maximum of the first harmonic can be estimated from
the first maximum of the Airy factor. We write
$\lambda_1=\hat{\omega}_1(1+\epsilon)$ \cite{malakhov2026analyticalcompton}, with
\begin{equation}
  \epsilon
  =
  |v_1|
  \left(
      \frac{1-\hat{\omega}_1}{\tau^2}
  \right)^{1/3}.
  \label{eq:app-first-maximum}
\end{equation}
Here $|v_1|\approx 1.0188$ is the absolute value of the Airy argument at the
first maximum of $\Ai(-v)$.

\section{NCS-Net Neural Network Architecture}
\label{appx:nn_arch}
\begin{figure}[b]
  \vspace{-20pt}
\includegraphics[width=\linewidth]{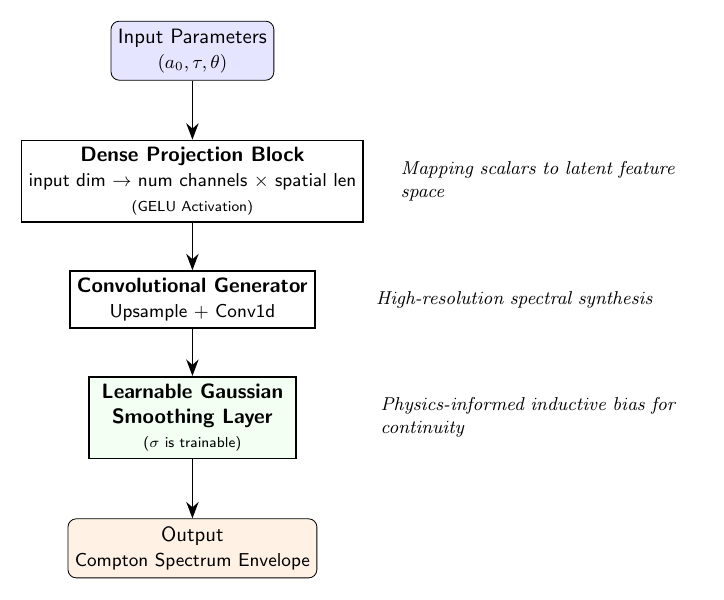}
\vspace{-20pt}
\caption{\label{fig:architecture} NCS-Net architecture.}
\end{figure}

NCS-Net neural network is divided into three functional blocks (Fig.~\ref{fig:architecture}). Initially, the input parameters are processed by the \texttt{Dense Projection Block} -- a cascade of fully connected layers with GELU activation functions. The vector dimension is sequentially expanded to form an initial tensor with shape (number of channels) $\times$ (spatial length). The dense block maps macroscopic interaction conditions into a latent feature space. This stage is responsible for encoding macroscopic spectral characteristics.

The transformation of the rough latent representation into the final high-resolution frequency domain is achieved through the \texttt{Convolutional Generator} -- several successive upsampling stages. In each stage, the spatial resolution is scaled using linear interpolation (\texttt{Upsample}), followed by 1D convolutions (\texttt{Conv1d}) with a wide receptive field, batch normalization, and GELU activation. The combination of linear interpolation and standard convolutions eliminates high-frequency oscillations known as checkerboard artifacts, which compromise physical fidelity, thereby ensuring smoothness. Wide convolution kernels allow the network to capture long-range spectral correlations.

Finally, the signal passes through a custom \texttt{Learnable Gaussian Smoothing Layer} (Gaussian convolution with trainable variance $\sigma$). This acts as a physics-informed inductive bias, guaranteeing function continuity and suppressing microscopic noise. Using a learnable $\sigma$ allows the model to autonomously determine optimal smoothing without artificially broadening spectral peaks. The model is implemented using the PyTorch framework.

The training of the NCS-Net surrogate neural network was performed using a composite multi-component loss function that integrates standard regression metrics with physics-informed regularizers. The total loss function $\mathcal{L}_{\text{total}}$ is defined as:
\vspace{-3pt}
\[
    \mathcal{L}_{\text{total}} = \mathcal{L}_{\text{main}} + \lambda_{\text{phys}} \left( \mathcal{L}_{\text{area}} + \mathcal{L}_{\text{peak}} + \mathcal{L}_{\text{ampl}} \right) + \lambda_{TV} \mathcal{L}_{TV},
    \vspace{-3pt}
\]
where $\mathcal{L}_{\text{main}}$ is the primary objective (Huber loss \cite{huber1992robust}) calculated in the normalized logarithmic space of intensities, $\lambda_{\text{phys}}$ and $\lambda_{TV}$ are weighting coefficients for the physics-informed constraints and smoothness regularizers.

To ensure that the model predictions correspond to the fundamental properties of Compton scattering, additional physics-informed components are computed in the linear intensity scale. The losses $\mathcal{L}_{\text{area}}$ and $\mathcal{L}_{\text{ampl}}$ minimize the relative errors in predicting the total photon yield and the spectral peak value. The component $\mathcal{L}_{\text{peak}}$ minimizes the relative error in predicted peak position. Since the standard \texttt{argmax} operation for identifying the spectral peak frequency is non-differentiable, we implement a Soft-Argmax approximation \cite{chapelle2010gradient,blondel2020learning}. We employ the Huber loss for physics-informed metrics too. To stabilize training for both linear and nonlinear spectrum samples, Huber loss is applied to values normalized by the ground truth targets.
To suppress numerical oscillations arising during the training process from conflicts between the primary objective ($\mathcal{L}_{\text{main}}$) and physics-informed metrics ($\mathcal{L}_{\text{area}},\, \mathcal{L}_{\text{ampl}},\, \mathcal{L}_{\text{peak}}$), a Total Variation penalty $\mathcal{L}_{\text{TV}}$ is introduced \cite{rudin1992nonlinear,condat2013direct}. This term effectively acts as an implicit low-pass filter, penalizing abrupt changes between adjacent spectral bins. Furthermore, to resolve optimization conflicts where physical constraints might hinder the initial learning of the spectral topology, we implement a warm-up strategy: $\lambda_{\text{phys}}$ is set to zero for the first several epochs and then linearly increased to its nominal value.

\end{document}